\documentclass[12pt]{amsart}
\usepackage[dvips]{color}
\usepackage{amsmath}
\usepackage{amsxtra}
\usepackage{amscd}
\usepackage{amsthm}
\usepackage{amsfonts}
\usepackage{amssymb}
\usepackage{eucal}
\usepackage{epsfig}
\usepackage{graphics}
\def\({\left(}
\def\){\right)}

\newcommand{\mub}{\mbox{\boldmath$\mu$}}

\newcommand{\betab}{\mbox{\boldmath$\beta$}}
\newcommand{\gammab}{\mbox{\boldmath$\gamma$}}
\newcommand{\nn}{\nonumber}
\newcommand{\bea}{\begin{eqnarray}}
\newcommand{\ena}{\end{eqnarray}}
\def\bel{\begin{eqnarray}}
\def\enl{\end{eqnarray}}
\newcommand{\be}{\begin{eqnarray*}}
\newcommand{\en}{\end{eqnarray*}}
\newcommand{\ba}{\begin{array}}
\newcommand{\ea}{\end{array}}

\newenvironment{tenumerate}{
  \begin{enumerate}
  
  }{\end{enumerate}}
\newcommand{\bi}{\begin{tenumerate}}
\newcommand{\ei}{\end{tenumerate}}
\newcommand{\isoto}[1][]%
{{\mathop{\buildrel{\sim}\over\longrightarrow}\limits_{#1}}}

\def\[{\left[}
\def\]{\right]}

\newcommand{\al}{\alpha}

\numberwithin{equation}{section}

\def\bi{\mathbf{i}}

\begin{document}
\begin{title}{Reflection relations and fermionic basis.}
\end{title}
\author{S.~Negro and  F.~Smirnov}

\address{SN
Laboratoire de Physique Th{\'e}orique et
Hautes Energies, Universit{\'e} Pierre et Marie Curie,
Tour 13, 4$^{\rm er}$ {\'e}tage, 4 Place Jussieu
75252 Paris Cedex 05, France
\newline
\centerline{$\mathrm{and}$}
\newline
\hspace{0.3cm}Dipartimento di Fisica, Universit\'a degli Studi di Torino
Open Space Dottorandi, via Pietro Giuria 1, 10124, Turin}\email{negro@to.infn.it \hspace{0.1cm};\hspace{0.1cm} negro@lpthe.jussieu.fr}

\address{
FS\footnote
{Membre du CNRS}Laboratoire de Physique Th{\'e}orique et
Hautes Energies, Universit{\'e} Pierre et Marie Curie,
Tour 13, 4$^{\rm er}$ {\'e}tage, 4 Place Jussieu
75252 Paris Cedex 05, France}\email{smirnov@lpthe.jussieu.fr}

\begin{abstract}
There are two approaches to computing the one-point functions for
sine-Gordon model in infinite volume. One is based on the use of the reflection relations,
this is a bootstrap type procedure. Another is based on using the fermionic basis which
originated in the study of lattice model. We show that the two procedures
are deeply interrelated.
 \end{abstract}

\maketitle

\section{Introduction}\label{intro}
In this paper we compare two approaches to the computation of the one-point
functions for sine(h)-Gordon model defined by the Euclidian action
\begin{align}
\mathcal{A}&=
\int \left\{ \Bigl[\frac 1 {4  \pi} \partial _z\varphi (z,\bar{z})\partial_{\bar{z}}\varphi (z,\bar{z})+
\frac{2\mub ^2}{\sin\pi b^2}\cosh(b\varphi(z,\bar{z}))
\right\}\frac{idz\wedge d\bar{z}}2\,.
\label{action1}
\end{align}
The  one-point functions in infinite volume  which
we shall deal with are given by the same formulae for the sinh-Gordon (shG) ($b$ is real)
and sine-Gordon (sG) ($b$ is pure imaginary) models \footnote{The normalisation of the 
dimensional constant is chosen because it happened to be very
convenient for the fermonic basis \cite{HGSV}, but there are serious physical
reasons for this choice. First, we have to change the sign of the potential energy going from the shG model to the sG model. For the latter the pole at
$b=i$ is natural because above this value the perturbation becomes irrelevant, and the usual treatment of the model breaks. The poles at integer $b$ for the shG model may look strange, but they become quite natural looking at
the formula for the physical scale of the model, i.e. for the mass of the particle
\cite{Alsinh}.
This point is the main reason for our normalisation because accepting it we have a universal formula for the mass $m$ of the shG particle and its close cousin,
the lowest breather in sG model \cite{Alsine}:
$$\mub \Gamma(1+b^2)=\[\frac {1} {4\sqrt{\pi}}\Gamma\Bigl(\frac {1}{2(1+b^2)}\Bigr)\Gamma\Bigl(\frac {2+3b^2}{2(1+b^2)}\Bigr)\cdot m
\]^{1+b^2}\,.$$
}.

The model can be considered as a perturbation of the $c=1$ Conformal Field Theory (CFT) by the relevant operator of dimension $\Delta=-b^2$. 
The idea which goes back to early days of the Perturbed Conformal Field Theory
(PCFT) \cite{RS, Alrestr} consists in another possible interpretation of the same model. Namely, rewriting the action as
\begin{align}
\mathcal{A}&=
\int \left\{ \Bigl[\frac 1 {4  \pi} \partial _z\varphi (z,\bar{z})\partial_{\bar{z}}\varphi (z,\bar{z})+
\frac{\mub ^2}{\sin\pi b^2} e^{b\varphi(z,\bar{z})}\Bigr]
+
\frac{\mub ^2}{\sin\pi b ^2}   e^{-b\varphi(z,\bar{z})}    \right\}\frac{idz\wedge d\bar{z}}2\,,
\label{action2}
\end{align}
we observe that the  model can be viewed as the perturbation of 
the Liouville model by the primary field of dimension $\Delta=-1-2b^2$. 
The Liouville model is not well-defined for imaginary $b$, for example, it
does not seem possible to write reasonable formulae for the three-point
functions. However, the ratio of the three-point functions of arbitrary descendants of the 
primary fields shifted by $mb$ ($m\in \mathbb{Z}$) to the three-point functions
of unshifted primary fields are defined, and this are the only data needed to
develop PCFT.

The PCFT allows to describe the ultra-violet behaviour of the perturbed model
using the ultraviolet data: Operator Product Expansion (OPE) for the perturbed
model,  and the infrared data: the one-point functions which depend
on the infrared environment. An excellent account of this procedure is given in the original paper \cite{Alpert} (see also
\cite{FFLZZ} for more details). Put the model  on the cylinder of radius $R$, which plays a role of the infrared cutoff. Then  the one-point functions are
subject to the perturbation theory
for small $R$. In the limit of infinite volume, $R\to\infty$, 
they approach certain non-perturbative values. 
So, the one-point functions provide an interesting example of the RG flow, and
their computation 
is an important problem.

Consider the primary field $e^{a\varphi(0)}$. In the present paper
we shall be interested in the one-point functions of its descendants. 
For the action \eqref{action1} they are naturally written as 
\begin{align}\frac {\langle P\bigl(\{\partial^k_z\varphi(0)\},\{\partial^k_{\bar z}\varphi(0)\}\bigr)e^{a\varphi(0)} \rangle} {\langle e^{a\varphi(0)} \rangle}\,,\label{ratio1}
\end{align}
where the normal ordering is implied which is derived from the T-ordering with respect
to the Euclidian time $t=1/2\log(z\bar{z})$. We shall consider  the polynomials $P$ of
{\it even} degree only. Then the invariance of \eqref{action1} under $\varphi\to-\varphi$
implies that \eqref{ratio1} is invariant
under
\begin{align}\sigma_1:\ a\to-a\,.\label{refl1}\end{align}

Now let us turn to \eqref{action2}. Introduce the components of the energy-mometum tensor
$$T_{z,z}(z,\bar z)=-\frac 1 4\partial_z\varphi(z,\bar z)^2+\frac Q 2\partial _z^2 \varphi(z,\bar z),\ \ 
T_{\bar z,\bar z}(z,\bar z)=-\frac 1 4\partial_{\bar z}\varphi(z,\bar z)^2+\frac Q 2\partial _{\bar z}^2 \varphi(z,\bar z)
\,,$$
here and later 
$$Q=b+\frac 1 b\,.$$
 We shall not use the trace of this energy-momentum tensor which is obviously proportional to $e^{-b\varphi(z,\bar z)}$.
Now the natural one-point functions are:
\begin{align}\frac {\langle L\bigl(\{\partial^k_zT_{z,z}(0)\},\{\partial^k_{\bar z}
T_{\bar z,\bar z}(0)\}\bigr)e^{a\varphi(0)} \rangle} {\langle e^{a\varphi(0)} \rangle}\,,\label{ratio2}
\end{align}
where $L$ is a polynomial.
One may assume that these one-point functions inherit the symmetry with 
respect to natural in the Liouville model reflection:
\begin{align}\sigma_2:\ a\to Q-a\,.\label{refl2}\end{align}

The idea of the paper \cite{FFLZZ} going back to 
\cite{FLZZ1,FLZZ2}  is to use the reflections
$\sigma_1,\sigma _2$ in order to compute the one-point functions. 
For arbitrary $a$ (in absence of resonances) the space of
local fields for the perturbed model is supposed to be the same as in corresponding CFT. One tries to apply this to both descriptions, namely
to use Heisenberg and Virasoro descendants.
 In sG case this is possible because they correspond to the Feigin-Fuchs 
bosonisation of the Virasoro algebra. In shG case following \cite{Seiberg,ZZ}
one can argue that both descriptions are available at the domain where
the Liouville zero-mode $\varphi_0$ goes to $-\infty$. 
Since the
two descriptions are related, one can start, for example, with Virasoro
descendants for which $\sigma _2$ is automatic. Then one rewrites them into
the Heisenberg descendants, which gives rise to a certain
matrix $U(a)$. This procedure is not completely trivial since one has to
factor out the action of the local integrals of motion as explained in Section \ref{reflection}.
The same is to be done for the second chirality.
Combining the expectation values \eqref{ratio2} into the vector
$V_{N,\bar{N}}(a)$, where $N$, $\bar{N}$ are multi indices counting descendants,
one finally arrives at the equations:
\begin{align}
V(Q-a)=V(a),\quad  V(a+Q)=(S(a)\otimes\bar{S}(a))V(a)\,,\label{RH}
 \end{align}
where 
$$S(a)=U(-a)U(a)^{-1}\,.$$
This Riemann-Hilbert problem was  introduced in \cite{FFLZZ}
where it is called reflection relations.
By analogy with the scattering theory, $S$ is a counterpart of $S$-matrix, and
$U(a)$ is a counterpart of wave operator, hence the notations.

The analyticity of  expectation values in infinite volume,
and the applicability of both reflections $\sigma _1$, $\sigma _2$ to
the perturbed model are rather subtle issues. This  is discussed
in details in the paper \cite{FLZZ2}.

The case of descendant operators, which the present paper concerns,  is considered in the paper \cite{FFLZZ} .  For level $2$ there
is only one nontrivial descendant, and the reflection relations can be solved.
There is a   CDD-like ambiguity, usual for bootstrap procedures, which 
is fixed by minimality assumption. There is also a problem of overall
normalisation which is ingeniously solved by requirement of cancellation of
resonances. All together a very reasonable expression for the one-point
function is obtained which is compared with the form factor expansion for
Lee-Yang model with impressive results. 
Going to higher levels is
problematic unless there are some independent hints about the structure
of solutions. We are going to describe now where these hints come from.

Another approach to computation of one-point functions 
proposed in \cite{OP,HGSV} uses the fermionic basis for counting the
descendants. The fermions acting on the space of local operators
for lattice six-vertex model (homogeneous or inhomonegeous) were
introduced in 
the papers
 \cite{HGS,HGSII,HGSIII}. The most important among these paper
is \cite{HGSIII} where using the fermionic description of local operators the expectation values for the six-vertex model  on
a cylinder are expressed
as determinants of one universal
function. This function allows description in terms of Thermodynamic Bethe Ansatz (TBA) \cite{BG}.
Thus a situation unique in the Quantum Field Theory  occurs  when the physically
relevant quantities (expectation values) are obtained in presence of the ultraviolet (lattice)
and infrared (cylinder of finite radius) cutoffs.

The logic of the papers  \cite{HGSIV,OP,HGSV} may be not very clear, let us streamline it.
The goal is to obtain the one-point functions for the sG model implying that the local
operators in the massive model can be identified with those in corresponding CFT.
The procedure consists to three steps.
\begin{itemize}
\item
Introducing the fermonic basis in the space of local operators for the lattice six-vertex model (homogeneous and inhomogeneous) \cite{HGS,HGSII,HGSIII}. This plays a role of
{\it Existence Theorem} for the fermonic basis in the continuous limit. 
\item
Taking the scaling limit for the homogeneous case we find the fermonic basis for
the CFT. This plays a role of {\it Normalisation}. 
\item 
Taking the scaling limit for the inhomogeneous case we find the one-point functions in
sG model for the operators which have being already identified. This is the {\it Goal}.
\end{itemize}
From that prospective the present paper contributes to better understanding and simplification
of the second item.

The notations used in  this paper are different from those of \cite{HGSIV,OP,HGSV}. The point
is that using intensively formulae from Liouville theory we decided to switch to the  notations of Zamolodchikov and Zamolodchikov \cite{ZZ}
following the  proverb {\it When in Rome, do as the Romans do.}
The identification with the notations of \cite{HGSIV,OP,HGSV} is 
$$b^2=\nu -1,\quad \nu\al =2ab\,.
$$

In fermionic description the descendants (modulo the action of local integrals of motion) are
created by two kind of operators $\betab^*_{2j-1},\gammab^*_{2j-1}$ applied in equal number.
Analysing the construction of the above papers one expects that when acting on the 
primary field $e^{a\varphi (0)}$ these operators transform as
\begin{align}\betab^*_{2j-1}\longrightarrow\gammab^*_{2j-1}
\qquad \gammab^*_{2j-1}\longrightarrow\betab^*_{2j-1}\,,\label{reflferm}
\end{align}
under both reflections $\sigma _1$ and $\sigma _2$. 
This is very transparent in the formulae for the one-point functions in infinite volume:
\begin{align}
&\frac
{\langle  
\bar{\betab} ^{*}_{\bar{I}^+}\bar{\gammab} ^{*}_{\bar{I}^-}
{\betab} ^{*}_{{I}^+}{\gammab} ^{*}_{{I}^-}e^{a\varphi (0)}
\rangle}
{\langle  
e^{a\varphi (0)}
\rangle}=\delta _{\bar{I}^-,I^+}\delta _{\bar{I}^+,I^-}
\(\frac{ 1}{1+b^2}\)^
{\#(I^+)+\#(I^-)}\label{main0}\\
&\times 
\mub^{\frac{2}{1+b^2}(|I^+|+|I^-|)}
\prod\limits _{2n-1\in I^+}
\tan\frac \pi {2Q} ((2n-1)b-2a)
\prod\limits _{2n-1\in I^-}
\tan\frac \pi {2Q} ((2n-1)b +2a)\,.
\nn
\end{align}
here and later we use the multiindex notations:
\begin{align}
&I=\{2i_1-1,\cdots ,2i_n-1\}\,,\quad |I|=\sum_{p=1}^n(2i_p-1)\,,\nn\\&
 \betab^*_I=\betab _{2i_1-1}\cdots \betab _{2i_n-1}\,,\quad
\gammab^*_I=\gammab _{2i_n-1}\cdots \gammab _{2i_1-1}\,.\nn
\end{align}
The formula \eqref{main0} may look too simple, but the relation between the fermionic descendants and the Virasoro ones is not simple. This explains how rather complicated formulae for the one-point
functions of the Virasoro descendants follow from simple formulae \eqref{main0}. 
Putting the same thing in other words the fermionic basis solves the reflection relations.
The main goal of the present paper consists in showing that the reflections \eqref{reflferm} are of purely algebraic nature.

The paper is organised as follows. In Section \ref{reflection} we define
the reflection matrix and give examples. In Section \ref{fermions} we explain the formulae 
relating the fermionic basis with the Virasoro one. The known formulae are up to level 8 \cite{HGSIV, Boos}. We explain that they indeed satisfy the reflection relations. In Section \ref{level10}
we inverse the logic. The reflection relations are complicated unless the structure of
solutions is not known in advance. But the existence of fermionic basis provides important
information about solutions. We explain how this can be used in order to find formulae relating
the fermionic basis with the Virasoro one
on the example of level 10. This procedure is much simpler than the one used in
\cite{HGSIV, Boos}.

\section{Reflection matrix}\label{reflection}

In order to
compute the normal ordered
expressions in CFT side one can use OPE, but it is more convenient to work in  the operator formalism.
In the conformal limit the field $\varphi(z,\bar z)$ splits into chiral parts: 
 $\varphi(z,\bar z)=\phi(z)+\phi (\bar z)$. We have the mode decomposition
 $$\phi(z)=\phi_0-2i\pi _0\log(z)+i\sum_{k\in \mathbb{Z}\backslash 0}^{\infty}\frac {a_k} kz^{-k}\,,$$ 
where the Heisenberg generators satisfy
$$[a_k,a_l]=2k\delta_{k,-l}\,,$$
and zero-mode is canonical:
$$ \pi_0=\frac{\partial}{i\partial\phi_0}\,.$$
All the formulae for the second chirality are similar, so, we shall not write them.

The primary field $e^{a\phi(0)}$ is identified with the highest vector of the Heisenberg algebra 
$$e^{a\phi(0)}
\Longleftrightarrow
e^{a\phi _0}|0\rangle\,,\quad a_k|0\rangle=0,\ \ k>0\,.$$
We shall mostly work with one chirality, but 
 in order to make the notations closer to those of \cite{HGSIV,HGSV}, we shall denote
$$\Phi _a= e^{a(\phi _0+\bar{\phi} _0)}|0\rangle\otimes\overline{|0\rangle}\,.$$
The second chirality will mostly remain a spectator.
Then the
correspondence with the local operators is:
$$
P\bigl(\{\partial^k_z\varphi(0)\}\}\bigr)e^{a\varphi(0)}
\Longleftrightarrow
P\bigl(\{i(k-1)!a_{-k}\}\}\bigr)\Phi _a\,.
$$
We shall consider only even polynomials $P$.

Now we introduce the generators of Virasoro algebra of central charge
$c=1+6Q^2$:
\begin{align}&\mathbf{l}_{k}=\frac 1 4\sum_{j\ne 0,k}a_{j}a_{k-j}+  (i(k+1)Q/2
+\pi _0)a_{k}\,,\quad k\ne 0\,,\label{vir}\\
&\mathbf{l}_{0}=\frac1 2\sum_{j=1}^{\infty}a_{-j}a_j+\pi_0(\pi_0+iQ)\,.\nn
\end{align}
$$\mathbf{l}_{0}\Phi _a=\Delta \Phi _a,\quad
\Delta=a(Q-a)\,.$$
The correspondence with the local operators is
$$L\bigl(\{\partial^k_zT_{z,z}(0)\}\bigr)e^{a\varphi(0)}
\Longleftrightarrow
L\bigl(\{k!\cdot\mathbf{l}_{-k-2}\}\bigr)\Phi _a\,.
$$

Additional restriction comes from the integrability. In CFT
we have an infinite series of integrals of motion: $\mathbf{i}_1,
\mathbf{i}_3, \mathbf{i}_5,\cdots$.
These integrals of motion survive the perturbation \cite{zamint}. We should not
count the descendants created by $\mathbf{i}_{2k-1}$'s because corresponding
one-point functions vanish. The easiest way to understand that is
to use the operator language in which the one-point function
is the average over the vacuum (eigenstate for integrals), and the action
of integrals on the local fields is given by commutators. 
On CFT side the factorisation is possible because the integrals of motion 
acting  on the Verma module $\mathcal{V}_a
$ respect both reflections $\sigma_1$ and $\sigma _2$.
So, we are
interested not in the entire Verma module  but in the quotient space:
\begin{align}
\mathcal{V}^\mathrm{quo}_a=
\mathcal{V}_a\ /\ \sum_{k=1}^\infty\mathbf{i}_{2k-1}\mathcal{V}_a\,.
\end{align}
We shall use $\equiv$ for equality in this space, i.e. equality modulo action of integrals 
of motion. Here are explicit formulae for the first three local integrals of motion which we shall need in this paper:
\begin{align}
&\mathbf{i}_1=\mathbf{l}_{-1}\,,\qquad\mathbf{i}_3=2\sum\limits_{k=-1}^{\infty}\mathbf{l}_{-3-k}\mathbf{l}_{k}\,,\nn\\
&\mathbf{i}_5=3\Bigl(\sum\limits_{k=-1}^{\infty}\sum\limits_{l=-1}^{\infty}
\mathbf{l}_{-5-k-l}\mathbf{l}_{l}\mathbf{l}_{k}
+\sum\limits_{k=-\infty}^{-2}\sum\limits_{l=-\infty}^{-2}
\mathbf{l}_{l}\mathbf{l}_{k}\mathbf{l}_{-5-k-l}\Bigr)
+\frac{c+2} 6
\sum\limits_{k=-1}^{\infty}(k+2)(k+3)\mathbf{l}_{-5-k}\mathbf{l}_{k}\,.\nn
\end{align}
Obviously the space $\mathcal{V}^\mathrm{quo}_a$ has nontrivial 
subspaces of even degrees $k$ only. We shall denote them by 
 $\mathcal{V}^\mathrm{quo}_{a,k}$. We have
 $$\mathrm{dim}\Bigl(\mathcal{V}^\mathrm{quo}_{a,k}\Bigr)=p(k/2)\,$$
 here and later $p(k/2)$ is the number of partitions of $k/2$.
Choose the basis ${v}^{(k)}_1,\cdots,{v}^{(k)}_{p(k/2)}$ in $ \mathcal{V}^\mathrm{quo}_{a,k}$ which is generated by lexicographically ordered
action of generators of Virasoro algebra with even indices. Let us take
also a basis ${h}^{(k)}_1,\cdots{h}^{(k)}_{p(k/2)}$ in the same space created
by action of even number of the Heisenberg algebra generators (examples
will be given later).
Then define first of all the matrix $U^{(k)}$:
\begin{align}{v}^{(k)}_i \equiv\sum_{j=1}^{p(k/2)}U^{(k)}_{i,j}(a){h}^{(k)}_j\,.\label{defU}\end{align}

Let us consider examples of constructing $U^{(k)}(a)$.
In order to simplify the notations we introduce the operators
$\mathbf{v}^{(k)}_i$,  $\mathbf{h}^{(k)}_i$ such that
$${v}^{(k)}_i=\mathbf{v}^{(k)}_i\Phi _a,\quad {h}^{(k)}_i=\mathbf{h}^{(k)}_i\Phi _a\,.$$

\vskip .2cm
\noindent{\it Level 2.}
\vskip .2cm
This case is rather trivial since the dimension equals one. 
Set
$$
\mathbf{v}^{(2)}_i =\mathbf{l}_{-2},\quad 
\mathbf{h}^{(2)}_j=(a_{-1})^2\,,
$$
and
obtain
\begin{align}{v}^{(2)}_1 \equiv\frac 1 4 (b+2a)(b^{-1}+2a){h}^{(2)}_1\,.\label{lev2}
\end{align}
Let us emphasise one more time that the vector $(\mathbf{l}_{-1})^2\Phi _a$
is dropped being a descendant of local integral of motion. 
The zeros of $(b+2a)(b^{-1}+2a)$ are singular vectors on level 2, which is not
surprising.

\vskip .2cm
\noindent{\it Level 4.}
\vskip .2cm
Set
\begin{align}
&\mathbf{v}^{(4)}_1=(\mathbf{l}_{-2})^2,\quad\mathbf{v}^{(4)}_2=
\mathbf{l}_{-4}\,;\qquad
\mathbf{h}^{(4)}_1=a_{-1}^4,\quad\ \ \ \mathbf{h}^{(4)}_2=a_{-2}^2\,.\nn
\end{align}
We have to understand which vectors should be factored out. First, these are
descendants of $\mathbf{i}_1$. On level four the integral $\mathbf{i}_3$ appear, but
we should not take it into account because its only descendant,
$\mathbf{i}_3\mathbf{i}_1\Phi _a$, is at the same time the
descendant of $\mathbf{i}_1$ due to commutativity. It is not hard to compute:
     \begin{align}
     \label{R4}
   U^{(4)}= -\frac1{144}\begin{pmatrix}
 -9 - 12 a^2 + 16 a^4 - 12 a Q + 8 a^3 Q, &  12 (3 + 16 a^2 + 14 a Q + 3 Q^2)\\
 4a  (2 a^3 + 3 a^2 Q  + 
   3 a ),& 12a(3 Q  + 2a )
   \end{pmatrix}  
     \end{align}
The determinant of this matrix is
$$
\det(U^{(4)})=C^{(4)}\cdot a (2 a + b)(1/b + 2 a ) (2 a + 3 b)  (3/b + 
   2 a ) (1/b + 2 a + b)\,,\nn
$$
where $C^{(4)}$ is an irrelevant numerical multiplier. Let us discuss this formula.
The multiplier $(2 a + b)(1/b + 2 a )$ corresponds to null-vectors coming from singular vectors on level 2. Its degree equals one because the descendants of
the local integrals of motion are factored out. The multiplier
$(2 a + 3 b)  (3/b + 
   2 a ) (1/b + 2 a + b)$ correspond to singular vectors appeared on level 4. The
   most peculiar multiplier is $a$. Obviously it should come from the
   singular vector existing for $a=0$ on the level 1: $\mathbf{l}_{-1}|0\rangle$.
   It may look strange that the singular vector from odd level has shown up in
   our construction, the explanation for this is given in the following line:
   $$
   0=\mathbf{l}_{-3}\mathbf{l}_{-1}|0\rangle=-2\mathbf{l}_{-4}|0\rangle+\mathbf{l}_{-1}\mathbf{l}_{-3}|0\rangle\equiv -2\mathbf{l}_{-4}|0\rangle\,.
   $$
  For higher levels the null-vectors obtained from the singular vectors at even levels
  come in regular way, while those created from the singular vectors at odd levels follow rather complicated pattern which we shall not explain since it is
  irrelevant for our goals. 
\vskip .2cm
\noindent{\it Level 6.}
\vskip .2cm
Starting from this case the formulae become rather heavy. 
So, we shall give only the most import ones. For the rest a Mathematica file
is available upon request. 
For the last time we define the Virasoro basis
$$\mathbf{v}^{(6)}_1=(\mathbf{l}_{-2})^3,\quad\mathbf{v}^{(6)}_2=\mathbf{l}_{-4}\mathbf{l}_{-2},\quad\mathbf{v}^{(6)}_3=\mathbf{l}_{-6}\,,$$
now our lexicographical rule is clear. We shall continue providing the Heisenberg bases because trying to repeat our computation one may be  unlucky enough to take vectors, linearly
dependent modulo integrals of motion. We set
$$\mathbf{h}^{(6)}_1=a_{-1}^6,\quad\mathbf{h}^{(6)}_2=a_{-1}^2a_{-2}^2,\quad\mathbf{h}^{(6)}_3=a_{-3}^2\,.$$
We factor out all the descendants of $\mathbf{i}_1$ and
one descendant of $\mathbf{i}_3$:
$$(\mathbf{i}_3)^2\Phi _a\,.$$
Another possible descendant of $\mathbf{i}_3$ being
$\mathbf{i}_3\mathbf{i}_1\mathbf{l}_{-2}\Phi _a$ is at the same time a descendant of $\mathbf{i}_1$, the same is true for the only possible
descendant of $\mathbf{i}_5$.

The formulae for the matrix elements of $U^{(6)}$ are complicated, but its determinant is given by nice and instructive formula:
\begin{align}\det(U^{(6)})=C^{(6)}\cdot N^{(6)}(a,b)\cdot\frac{\Delta + 2}{3 a^2-10 Q^2 - 5}\,,
\label{det6}
\end{align}
where
the null-vector contribution 
\begin{align}
N^{(6)}(a,b)&=a (2 a + b)^2 (2 a + 3 b) (2 a + 5 b) (1/b + 2 a )^2 (3/b + 
         2 a ) (5/b + 2 a ) \nn\\&\times(1/b + 2 a + b) (2/b + 2 a + 
         b) (1/b + 2 a + 2 b)\,,\nn
\end{align}
is not interesting for us. The remaining multipliers
have the following explanation.
First,
$$U^{(6)}=U^{(6)}_0+\frac 1 {3 a^2-10 Q^2 - 5} U^{(6)}_1\,$$
where the matrix $U^{(6)}_1$ depends linearly on $a$ and has rank 1. 

Second,
$$\bigl(U^{(6)}\bigr)^{-1}=\frac  1{\Delta+2}\ U_3^{(6)}+U_4^{(6)}\,,
$$
where $U_4^{(6)}$ is regular at $\Delta=-2$, and $U_3^{(6)}$ has rank 1. 
The coimage of $U_3^{(6)}$ will be rather important, it is span by
\begin{align}
\mathbf{w}^{(6)}= \mathbf{l}_{-4}\mathbf{l}_{-2}   + \frac{c-16} 2\mathbf{l}_{-6}\label{w6}
\,.
\end{align}
\vskip .2cm
\noindent{\it Level 8.}
\vskip .2cm
The Heisenberg basis is
\begin{align}
&\mathbf{h}^{(8)}_1=a_{-1}^8,\quad\mathbf{h}^{(8)}_2=a_{-1}^4a_{-2}^2,\quad\mathbf{h}^{(8)}_3=a_{-2}^4,%\nn\\&
\quad\mathbf{h}^{(8)}_4=a_{-4}^2,\quad\mathbf{h}^{(8)}_5=a_{-2}a_{-6}\,.\nn
\end{align}
We factor out all the descendants of $\mathbf{i}_1$ and two descendants of 
$\mathbf{i}_3$:
$$\mathbf{i}_3\mathbf{l}_{-5}\Phi _a,\ \ \mathbf{i}_3\mathbf{l}_{-3}
\mathbf{l}_{-2}\Phi _a\,,$$
then the descendants of $\mathbf{i}_5,\mathbf{i}_7$ do not count. 

The determinant of $U^{(8)} $ is given by
\begin{align}\det(U^{(8)})=C^{(8)}\cdot N^{(8)}(a,b)\cdot
\frac{(\Delta + 
   11) (\Delta + 4)}{a^2(-21 (76 - 19 Q^2 - 30 Q^4)- (991 + 1076 Q^2) a^2 + 206 a^4)}\,,\label{det8}
\end{align}

  where
  \begin{align}
&N^{(8)}(a,b)= a^2 (b + a)(1/b + 
   a) (b + 2 a)^3(1/b + 
   2 a)^3 (3 b + 2 a)^2 (3/b + 2 a)^2 \nn\\&\times(5 b + 2 a)(5/b + 2 a) (7 b + 2 a) (7/b + 2 a)(b + 1/b + 2 a)^2     \nn\\&\times (b + 3/b + 2 a) (3 b + 1/b + 2 a)(b + 2/b + 
   2 a) (2 b + 1/b + 2 a)\,.\nn
  \end{align}
  Notice the multiplier $(b + a)(1/b + 
   a)$ which signifies that  descendants of the singular vectors on level 3 contributed for the first time. 
  We have
$$ U^{(8)}=U^{(8)}_0+\frac 1 { a^2} U^{(8)}_2 +\frac 1 {21 (76 - 19 Q^2 - 30 Q^4)- (991 + 1076 Q^2) a^2 + 206 a^4} U^{(8)}_3\,,$$
where the ranks of $U^{(8)}_2$ and $U^{(8)}_3$ are respectively 1 and 2. 
More importantly for us
$$ \bigl(U^{(8)}\bigr)^{-1}=\frac 1 {\Delta+4} U^{(8)}_4 +\frac 1 {\Delta +11} U^{(8)}_5+U^{(8)}_6\,,$$
where the ranks of the matrices $U^{(8)}_4$ and $U^{(8)}_5$ equal 1, their coimages are span by 
\begin{align}
&\mathbf{w}^{(8)}_4=-28\ \mathbf{l}_{-4}(\mathbf{l}_{-2})^2 + 3 ( c-36) (\mathbf{l}_{-4})^2 -2( 5 c -12)\mathbf{l}_{-6}\mathbf{l}_{-2}  \label{w4w11}\\
&\qquad\quad +( 4128 - 325 c  + 5 c^2 )\mathbf{l}_{-8},\nn\\
&\mathbf{w}^{(8)}_{11}=3 (\mathbf{l}_{-4})^2 + 4 \mathbf{l}_{-6}\mathbf{l}_{-2} + (5 c-89)\mathbf{l}_{-8}\,.\nn
\end{align}
\vskip .2cm
\section{Solving reflection relations by fermionic basis}\label{fermions}

The main statement of the paper \cite{HGSIV} is that changing the normalisation of  fermions $\betab^*$, $\gammab^*$ one obtains purely CFT objects. Namely,
define $\betab^{\mathrm{CFT}*}_{2m-1}$, $\gammab^{\mathrm{CFT}*}_{2m-1}$ acting on $\mathcal{V}^{\mathrm{quo}}_a$ by
$$\betab^*_{2m-1}=D_{2m-1}(a)\betab^{\mathrm{CFT}*}_{2m-1},\quad
\gammab^*_{2m-1}=D_{2m-1}(Q-a)\gammab^{\mathrm{CFT}*}_{2m-1}\,,$$
where
$$D_{2m-1}(a)=(-1)^m\sqrt{\frac 1{i(1+b^2)}}\ \Gamma (1+b^2)^{-\frac {2m-1}{1+b^2}}b^{2m-1}
\frac{\Gamma \(\frac 1 {2Q}\( 2a  +(2m-1)b^{-1}\)\)}{(m-1)!\Gamma \(\frac 1 {2Q}\(2 a  - (2m-1){b} \)\)}\,.$$
Then for $\#(I^+)=\#(I^-)$ we have
\begin{align}
\betab_{I^+}^{\mathrm{CFT}*}\gammab_{I^-}^{\mathrm{CFT}*}\Phi _a\equiv C_{I^+,I^-}\cdot\Bigl(
P^{\mathrm{even}}_{I^+,I^-}(\{\mathbf{l}_{-2k}\},\Delta,c)+d\cdot
P^{\mathrm{odd}}_{I^+,I^-}(\{\mathbf{l}_{-2k}\},\Delta,c)
\Bigr)\Phi _a\,,\label{polP}
\end{align}
where 
$P^{\mathrm{even}}_{I^+,I^-}(\{\mathbf{l}_{-2k}\},\Delta,c)$, $P^{\mathrm{odd}}_{I^+,I^-}(\{\mathbf{l}_{-2k}\},\Delta,c)$ are polynomials in Virasoro generators 
with coefficients depending rationally on $\Delta$ and $c$,
there is one more constant\footnote{The sign of
$d$ is changed comparing to \cite{HGSIV}.}:
$$d={\textstyle\frac 1 6}\sqrt{(c-25)(24\Delta +1-c)}=(b-b^{-1})(Q-2a)\,,$$
$C_{I^+,I^-}$ is the Cauchy determinant:
$$C_{I^+,I^-}=\det \(\frac 1 {i^+_p+i^-_q-1}\)_{p,q=1,\cdots ,\#(I^+)}\,.
$$
We have
$$P^{\mathrm{even}}_{I^+,I^-}=P^{\mathrm{even}}_{I^-,I^+}\,,\quad
P^{\mathrm{odd}}_{I^+,I^-}=-P^{\mathrm{odd}}_{I^-,I^+}\,,$$
in particular
$$P^{\mathrm{odd}}_{I^+,I^-}=0\quad \mathrm{for}\quad I^+=I^-\,.$$
The Cauchy determinant was introduced in order to fix the normalisation (see \cite{HGSIV}):
$$P^{\mathrm{even}}_{I^+,I^-}(\{\mathbf{l}_{-2k},\Delta,c\})=
(\mathbf{l}_{-2})^{\frac 1 2(|I^+|+|I^-|)}+\cdots\,,$$
Later we shall provide examples. 

Let us discuss the reflections $\sigma _{1,2}$ for  fermions
$\betab^{\mathrm{CFT}*}_{2m-1}$, $\gammab^{\mathrm{CFT}*}_{2m-1}$ . Obviously under $\sigma _2$ they transform as original ones:
\begin{align}
\betab^{\mathrm{CFT}*}_{2m-1}\longrightarrow\gammab^{\mathrm{CFT}*}_{2m-1},\quad \gammab^{\mathrm{CFT}*}_{2m-1}\longrightarrow\betab^{\mathrm{CFT}*}_{2m-1}\,.\label{CFTfermionsrefl1}
\end{align}
But for the reflection $\sigma_1$ the transformation law changes because of 
$D_{2m-1}(a)$, $D_{2m-1}(Q-a)$:
\begin{align}
&\gammab^{\mathrm{CFT}*}_{2m-1}\longrightarrow\(\frac{2a  -(2m-1)b}{2a  +(2m-1)b^{-1}}\)\betab^{\mathrm{CFT}*}_{2m-1}\,,
\label{CFTfermionsrefl2}\\
&\betab^{\mathrm{CFT}*}_{2m-1}\longrightarrow\(\frac{2a  -(2m-1)b^{-1}}{2a  +(2m-1)b}\)\gammab^{\mathrm{CFT}*}_{2m-1}\,.\nn
\end{align}
The only way to satisfy this reflection is to assume that together with
\eqref{polP} we have another kind of formulae:
\begin{align}
&\betab^{\mathrm{CFT}*}_{I^+}\gammab^{\mathrm{CFT}*}_{I^-}\Phi_a\equiv C_{I^+,I^-}\cdot\prod_{2j-1\in I^+}(2a+(2j-1)b^{-1})
\prod_{2j-1\in I^-}(2a+(2j-1)b)\label{polQ}
\\&\times\Bigl(Q_{I^+,I^-}^{\mathrm{even}}(\{a_{-k}\},a^2,Q^2)+
g\cdot Q_{I^+,I^-}^{\mathrm{odd}}(\{a_{-k}\},a^2,Q^2)\Bigr)\Phi_a\,,\nn
\end{align}
where
$Q_{I^+,I^-}^{\mathrm{even}}(\{a_{-k}\},a^2,Q^2)$,
$Q_{I^+,I^-}^{\mathrm{odd}}(\{a_{-k}\},a^2,Q^2)$ are polynomials in Heisenberg generators depending rationally on $a^2$ and $Q^2$,
$$g=a(b-b^{-1})\,.$$
This is a very serious statement which has to be checked.
\vskip .2cm
\noindent
{\it Level 2.}
We have $P^{\mathrm{even}}_{\{1,1\}}=\mathbf{l}_{-2}$, 
$P^{\mathrm{odd}}_{\{1,1\}}=0$ (from now on we shall not write vanishing by definition polynomials). Then the formula \eqref{lev2} shows that
$Q^{\mathrm{even}}_{\{1,1\}}=\frac 1 4 (a_{-1})^2\,.$
\vskip.2cm

\noindent
{\it Level 4.}
\vskip.2cm
We have the following formulae form \cite{HGSIV}:
$$P_{\{1\},\{3\}}^{\mathrm{even}}=(\mathbf{l}_{-2} )^2+ \frac{2 c - 32}{9}\ \mathbf{l}_{-4} ,\quad
P_{\{1\},\{3\}}^{\mathrm{odd}}= \frac2 3\ \mathbf{l}_{-4}\,.$$
Applying the matrix $U^{(4)}$ \eqref{R4} one observes that in the
expression $P_{\{1\},\{3\}}^{\mathrm{even}}+dP_{\{1\},\{3\}}^{\mathrm{odd}}$
the multiplier $(2a+b^{-1})(2a +3b)$ factorises leaving
\begin{align}
&Q_{\{1,3\}}^{\mathrm{even}}=-\frac1{144 } \Bigl\{(4 a^2 (Q^2 - 2)-3) (a_{-1})^4 + 
   12(1+Q^2) a_{-2}^2 \Bigr\} \nn\\
&Q_{\{1,3\}}^{\mathrm{odd}}=\frac 1 {216} \Bigl\{(-3 + 4 a^2) (a_{-1})^4 + 12 a_{-2}^2)\Bigr\}\,.\nn
\end{align}

\noindent
{\it Level 6.}
\vskip.2cm
The polynomials $P^{\mathrm{even}\atop\mathrm{odd}}$ for level 6 were computed in \cite{HGSIV}.
It came as a surprise that the coefficients contain the denominator $\Delta+2$. 
Now it follows clearly from the formula for the determinant \eqref{det6}. Moreover, it is clear
that  for all the polynomials the residues at $\Delta =-2$ are proportional to the vector $\mathbf{w}^{(6)}$
\eqref{w6}.
This allows to simplify the formulae of \cite{HGSIV}:
\begin{align}
P_{\{3\},\{3\}}^{\mathrm{even}}(\{\mathbf{l}_{-2k}\})&= (\mathbf{l}_{-2})^3  + 
   \frac 2 3 (c-19 ) \mathbf{l}_{-4}\mathbf{l}_{-2} + \frac 1 {30}(1524 - 173 c + 5 c^2  + 
      8 (c-28) \Delta)\mathbf{l}_{-6}\nn\\
     & -\frac1{6(\Delta+2)} (  5 c-158) \mathbf{w}^{(6)}\,,\nn\\
P_{\{1\},\{5\}}^{\mathrm{even}}(\{\mathbf{l}_{-2k}\})&=     (\mathbf{l}_{-2})^3   
   +\frac 2 3 ( c-10)\mathbf{l}_{-4}\mathbf{l}_{-2}  + \frac 1 {15}(140 - 59 c + 3 c^2  +
      8 (c-28) \Delta) \mathbf{l}_{-6}\nn\\
      & -\frac1{(\Delta+2)} (  c-14) \mathbf{w}^{(6)}\,,\nn\\
    P_{\{1\},\{5\}}^{\mathrm{odd}}(\{\mathbf{l}_{-2k}\})&=    2  \mathbf{l}_{-4}\mathbf{l}_{-2} +\frac 4 {5} (c-13 ) \mathbf{l}_{-6}
  -\frac 4{(\Delta+2)} \mathbf{w}^{(6)}\,.\nn  
\end{align}
Now using the matrix $ U^{(6)}(a)$ we make sure that the factorisation  \eqref{polQ} takes place.
We give the formulae for $Q^{\mathrm{even}}_{\{1\},\{5\}}$, $Q^{\mathrm{odd}}_{\{1\},\{5\}}$, $Q^{\mathrm{even}}_{\{3\},\{3\}}$ in the Appendix. One can check directly, without computing $U^{(6)}(a)$ that
the differences between $P$ and $Q$ expressions are linear combinations of descendants 
of integrals of motion.

\noindent
{\it Level 8.}
\vskip.2cm

The formulae for the fermionic basis on the level 8 are interesting
because this is the first time when a state containing four fermions
appear. 
However, in order to make the computations  following the procedure
of \cite{HGSIV} one is forced to consider descendants in the Matsubara direction.
This is a hard work which was done by H. Boos \cite{Boos}. 
He observed that there are two denominators $\Delta +4$ and $\Delta +11$.
Once again, this is not surprising having in mind \eqref{det8}. Corresponding
residues are proportional to $\mathbf{w}^{(8)}_4$, $\mathbf{w}^{(8)}_{11}$ \eqref{w4w11}.
We give only the most interesting example for the state with four fermions, the
rest can be found in \cite{Boos}:
\begin{align}
&P_{\{1,3\},\{1,3\}}^{\mathrm{even}}(\{\mathbf{l}_{-2k}\})=
(\mathbf{l}_{-2} )^4+\frac{4(c-22)}3\mathbf{l}_{-4}(\mathbf{l}_{-2})^2
-\frac 1 {9} (c^2- 34 c-333   + 8 (c-25) \Delta)(\mathbf{l}_{-4})^2
\nn\\&+\frac 2 {15} (  5 c^2 - 193 c +1544+ 8 (c -28)\Delta)\mathbf{l}_{-6}\mathbf{l}_{-2}\nn\\&-\frac 4 3 (11 c-71  + 24 \Delta)\mathbf{l}_{-8}
+\frac {5c-122} {42(\Delta +4)}\mathbf{w}^{(8)}_4
- \frac{8648 - 526 c + 5 c^2}{42(\Delta +11)} \mathbf{w}^{(8)}_{11}\,.\nn
\end{align}
It gives a great satisfaction to observe how the multiplier $(2a+b)(2a+3b)(2a+b^{-1})(2a+3b^{-1})$
factors when the Virasoro basis is changed to the Heisenberg one, leaving an even in $a$ function
$Q^{\mathrm{even}}_{\{1,3\},\{1,3\}}$. The latter function is presented in the Appendix. 
We checked the formula \eqref{polQ} for all other  vectors given in \cite{Boos}. 
We consider this as an independent check of them.

Suppose we have a linear functional $f$ on $\mathcal{V}_a^{\mathrm{quo}}$
such that the vectors
$$V_i(a)=f(\mathbf{v}_i\Phi_a)\,,\qquad H_i(a)=f(\mathbf{h}_i\Phi_a)\,,$$
satisfy the  relations
\begin{align} 
V(Q-a)=V(a),\qquad H(-a)=H(a)\,.\label{VH}
\end{align}
These relations  together with $V(a)=U(a)H(a)$ imply the nontrivial
Riemann-Hilbert problem:
\begin{align}V(a+Q)=S(a)V(a)\,,\qquad S(a)=U(-a)U(a)^{-1}\,,\label{RHchiral}
\end{align}
which is the chiral part of  \eqref{RH}.
Introduce the vector
$$W_{I^+,I^-}(a)=f(\betab^*_{I^+}\gammab^*_{I^-}\Phi _a)\,.$$
The results of this section are summarised by two equations:
\begin{align}
W(-a)=JW(a)\,,\qquad W(Q-a)=JW(a)\,,\label{W=JW}
\end{align}
where the matrix $J$ interchanges the components $I^+,I^-$ and $I^-,I^+$.
 Going to the fermonic basis we
managed to transform the nontrivial Riemann-Hilbert problem (\ref{VH},\ref{RHchiral})  to the trivial  one \eqref{W=JW}.
Oppositely, taking any solution of \eqref{W=JW}, and performing the change of basis  inverse 
to \eqref{polP}
one gets a solution to (\ref{VH}), (\ref{RHchiral}). In this way we obtain a complete set of solutions
which can be combined with the quasi-constant coefficient (scalar functions
satisfying $g(-a)=g(a)$, $g(Q-a)=g(a)$).

Reflection relations themselves do not 
provide an unique way of gluing two chiralities for the one-point functions in infinite
volume. The formulae from \cite{OP} shows that the correct way is as follows.
Going to the second chirality we make the change $a\to Q-a$ \cite{OP,HGSV}. Define $\overline{W}(a)$ similarly to $W(a)$. Then the one-point functions
correspond to the choice of $W(a)\times\overline{W}(a)$ cited in the Introduction \eqref{main0}.

\section{Determining fermionic basis from reflection}\label{level10}

The procedure of determining the fermionic basis from the determinant formula
described in \cite{HGSIV,Boos} becomes very complicated starting from the level 8.
Let us show that the reflection gives much simpler way provided the {\it a priori} knowledge 
of the fermonic basis exists.

Consider level $k$. First, one constructs the matrix $U^{(k)}(a)$. Its matrix elements
have the denominator $D^{(k)}_H(a^2,Q^2)$, for example,
$$D^{(6)}_H(a^2)=5 - 
      3 a^2 + 10 Q^2\,.$$
      We need to determine the Virasoro denominator. To this end we compute the
      determinant:
\begin{align}\det(U^{(k)})=C^{(k)}\cdot N^{(k)}(a,b)\cdot\frac{D_V^{(k)}(\Delta,c)}{D^{(k)}_H(a^2,Q^2)}\,,
\label{detk}
\end{align}     
We look for $P^{\mathrm{even}\atop\mathrm{odd}}_{I^+,I^-}$ in the form
\begin{align}
&P^{\mathrm{even}}_{I^+,I^-}=\mathbf{v}_1+\frac1 {D_V^{(k)}(\Delta,c)}\sum_{i=2}^{p(k/2)}X_{I^+,I^-,i}(\Delta,c)\mathbf{v}_i\,,\nn\\
&P^{\mathrm{odd}}_{I^+,I^-}=\frac1 {D_V^{(k)}(\Delta,c)}\sum_{i=2}^{p(k/2)}Y_{I^+,I^-,i}(\Delta,c)\mathbf{v}_i\,,\nn
\end{align}
where $\mathbf{v}_i$ are lexicographical as usual, 
$X_{I^+,I^-,i}(\Delta,c)$, $Y_{I^+,I^-,i}(\Delta,c)$ are
polynomials in $\Delta$ of degree $D$. We do not specify $D$ for the moment. We consider the coefficients of these polynomials as
unknowns, there are 
\begin{align}\#(\mathrm{unknowns})=2(p(k/2)-1)\cdot (D+1)\,,\label{unkn}
\end{align} of them. 

Let us introduce the polynomials
\begin{align}&T_{I^+,I^-}^{+}(a)= \frac 1 2\Bigl\{\prod _{2j-1\in I^+}(2a+(2j-1)b^{-1})
\prod _{2j-1\in I^-}(2a+(2j-1)b)\nn\\&\qquad\qquad+\prod _{2j-1\in I^+}(2a+(2j-1)b)
\prod _{2j-1\in I^-}(2a+(2j-1)b^{-1})\Bigr\}\,,\nn\\
&T_{I^+,I^-}^{-}(a)= \frac 1 {2a(b-b^{-1})}\Bigl\{\prod _{2j-1\in I^+}(2a+(2j-1)b^{-1})
\prod _{2j-1\in I^-}(2a+(2j-1)b)\nn\\&\qquad\qquad-\prod _{2j-1\in I^+}(2a+(2j-1)b)
\prod _{2j-1\in I^-}(2a+(2j-1)b^{-1})\Bigr\}\,.\nn
\end{align}
These polynomials are invariant under $b\to b^{-1}$, hence they depend on $b$ only
through $Q$.

Now it  is easy to see that the equations \eqref{polQ} are equivalent to two 
polynomial requirement which hold for any $1\le j\le p(k/2)$.
\newline
\noindent
First, 
\begin{align}
&D_V^{(k)}(\Delta(-a),c)D_H^{(k)}(a^2,Q^2)\label{req1}\\&\times\Bigl\{T_{I^+,I^-}^{+}(-a)\Bigl(D_V^{(k)}(\Delta,c)U_{1,j}^{(k)}(a)+
\sum_{i=2}^{p(k/2)}X_{I^+,I^-,i}(\Delta,c)U_{i,j}^{(k)}(a)\Bigr)\nn\\&-(Q^2-4)(Q-2a)T_{I^+,I^-}^{-}(-a)\sum_{i=2}^{p(k/2)}Y_{I^+,I^-,i}(\Delta,c)U^{(k)}_{i,j}(a)\Bigr\}\quad\mathrm{is\ even\ in\ }a\,,\nn
\end{align}
Second, 
\begin{align}
&D_V^{(k)}(\Delta(-a),c)D^{(k)}_H(a^2,Q^2)\label{req2}\\&\times\Bigl\{-T_{I^+,I^-}^{-}(-a)\Bigl(D_V^{(k)}(\Delta,c)U_{1,j}^{(k)}(a)+\sum_{i=2}^{p(k/2)}X_{I^+,I^-,i}(\Delta,c)U_{i,j}^{(k)}(a)\Bigr)\nn\\&+(Q-2a)T_{I^+,I^-}^{+}(-a)\sum_{i=2}^{p(k/2)}Y_{I^+,I^-,i}(\Delta,c)U_{i,j}^{(k)}(a)\Bigr\}\quad\mathrm{is\ odd\ in\ }a\,,\nn
\end{align}
These requirement are linear equations for our unknowns. 
We have
\begin{align}
&\#(\mathrm{equations})\label{eqs}\\&=\bigl(2\mathrm{deg}_{\Delta}(D_V^{(k)}(\Delta,c))+\mathrm{deg}_{a}(D_H^{(k)}(a^2,Q^2)U^{(k)}(a))
+2\#(I^+)+2D+1\bigr)\cdot p(k)\,,\nn
\end{align}
of them. Thus the system is overdetermened, and the very existence of solution is a miracle
produced by our fermionic basis. We considered as an example the case of level 10. 

For level 10 we take the following basis on the Heisenberg side:
\begin{align}
&\mathbf{h}_1^{(10)}=({a}_{-1})^{10},\ \mathbf{h}_2^{(10)}=({a}_{-1})^{2}({a}_{-2})^{4},\ \mathbf{h}_3^{(10)}=({a}_{-2})^{2}({a}_{-3})^{2},\ \mathbf{h}_4^{(10)}=({a}_{-1})^{5}{a}_{-5}, \nn\\& \mathbf{h}_5^{(10)}=({a}_{-5})^2,
\ \mathbf{h}_6^{(10)}=({a}_{-1})^{3}{a}_{-7}, \ \mathbf{h}_7^{(10)}={a}_{-1}{a}_{-9}\,.\nn
\end{align}
We compute the matrix $U^{(10)}(a)$, finding in particular:
\begin{align}
&D_V^{(10)}(\Delta,c)=\Bigl( \Delta +6  \Bigr)\nn\\&\times\Bigl(-23794 + 
          2905 c + (-2285 + 983 c)\Delta + (1447 + 71 c)
           \Delta^2 + (149 + c)\Delta^3 + 3 \Delta^4\Bigr)\,,\nn\\
&D_H^{(10)}  (a^2,Q^2)=a^2\Bigl(  % \nn\\&
          2025 (6 + 19 Q^2 + 16 Q^4 + 4 Q^6)  \nn\\&+ 
          5 a^2 (-5701 - 5153 Q^2 + 2793 Q^4 + 5562 Q^6 + 1944 Q^8)\nn\\& - 
          a^4 (53317 + 72222 Q^2 + 67739 Q^4 + 28326 Q^6)
          +a^6 (10657 + 21920 Q^2 + 27282 Q^4)\nn\\&+ 
          a^8 (11097 - 9810 Q^2)+1134 a^{10} \Bigr)  \,.\nn       
\end{align}
No contributions from  new  singular vectors of odd level appear in the multiplier
$ N^{(10)}(a,b)$ comparing to $ N^{(8)}(a,b)$. So, the contribution from
these singular vectors is $a^3(a+b)(a+b^{-1})$. The contributions of the singular
vectors of even level follow the usual routine. 

Now we can apply the procedure described above to find the fermionic basis. We start with 
$D=9$ which give a comfortable margin. The equations allow solutions for all possible
cases: $\{1\},\{9\}$; $\{3\},\{7\}$; $\{5\},\{5\}$ and $\{1,3\},\{1,5\}$. The actual degrees $D$ are
\begin{align}
&D=7:\quad \{1\},\{9\}\ \mathrm{even},\ \{3\},\{7\}\ \mathrm{even},\ \{5\},\{5\}\ \mathrm{even},\nn\\
&D=6:\quad \{1\},\{9\}\ \mathrm{odd},\ \ \{3\},\{7\}\ \mathrm{odd},\ \ \{1,3\},\{1,5\}\ \mathrm{even,\ odd}\,.\nn
\end{align}
The general structure of the answers is similar to what we had before. There is one vector
$\mathbf{w}^{(10)}_{6}$ which corresponds to the multiplier $\Delta +6$ in $D_V^{(10)}(\Delta,c)$,
there are four vectors $\mathbf{w}^{(10)}_{\mathrm{deg}4,i}$, $i=1,\cdots, 4$ corresponding
to the polynomial of degree 4 in $D_V^{(10)}(\Delta,c)$. The formulae are rather hard we shall not cite them, they are available upon request. Our goal was  to check that the computation using
the requirements \eqref{req1}, \eqref{req2} is possible. 
Also the experimental data should be useful for solving
the main remaining problem which consists
in constructing OPE for PCFT in the fermionic basis. 
Solving this problem is important for ameliorating the formulae for ultra-violet
asymptotics of two-point correlation function.

%%%%%%%%%%%%%%%%%%%%%%
%%%%%%%%%%%%%%%%%%%%%%%
%%%%%%%%%%%%%%%%%%%%%%%
\section{Appendix}
This Appendix contains formulae for $Q^{\mathrm{even}}$, $Q^{\mathrm{odd}}$
on levels 6 and 8.
\begin{align}
Q_{\{3\},\{3\}}^{\mathrm{even}}(\{a_{-k}\})&=\frac1{129600}\Bigl\{-\bigl[720 a^4 (3 + 2 Q^2) + 12 a^2 (18 + 341 Q^2 + 70 Q^4) \nn\\&+ 
      5 (771 + 2876 Q^2 + 2768 Q^4 + 560 Q^6)\bigr] a_{-1}^6\nn\\& - 
  1800 \bigl[18 + 32 Q^2 + 7 Q^4 + 12 a^2 (3 + Q^2)\bigr] a_{-1}^2 a_{-2}^2 \nn\\&+ 
  240 \bigl[138 + 293 Q^2 + 94 Q^4 - 12 a^2 (-9 + 2 Q^2)\bigr] a_{-3}^2) \nn\\&
  +50  \frac{(2 Q^2 + 1)^2 (14 Q^2 + 51)}{5 - 
      3 a^2 + 10 Q^2}\mathbf{g}\Bigr\}\,,\nn
\end{align}
\begin{align}
Q_{\{1\},\{5\}}^{\mathrm{even}}(\{a_{-k}\})&=\frac1{129600}\Bigl\{\bigl[-1440 a^4 (-1 + 2 Q^2) + a^2 (5412 + 1944 Q^2 - 4080 Q^4)\nn\\& + 
     5 \bigl[1009 + 3080 Q^2 + 224 Q^4 - 2720 Q^6)\bigr] a_{-1}^6\nn\\& - 
  900 \bigl[-11 - 51 Q^2 + 68 Q^4 + 12 a^2 (1 + 4 Q^2)\bigr] a_{-1}^2 a_{-2}^2\nn\\& + 
  480 \bigl[-8 - 33 Q^2 + 146 Q^4 + 12 a^2 (-9 + 2 Q^2)\bigr] a_{-3}^2\nn\\&+100\frac{(1 + 2 Q^2) (-29 - 60 Q^2 + 68 Q^4)}{5 - 
      3 a^2 + 10 Q^2} \mathbf{g}\Bigr\}\,,\nn
\end{align}
\begin{align}
Q_{\{1\},\{5\}}^{\mathrm{odd}}(\{a_{-k}\})&=\frac 1 {32400} \Bigl\{\bigl[2063 + 360 a^4 + 4216 Q^2 + 1920 Q^4 + 
      30 a^2 (34 + 21 Q^2)\bigr] a_{-1}^6 \nn\\&+ 
   450 \bigl[34 + 12 a^2 + 21 Q^2\bigr] a_{-1}^2 a_{-2}^2 - 
   3840 \bigl[4 + 3 Q^2\bigr] a_{-3}^2
\nn\\&+10\frac{ (1 + 2 Q^2) (67 + 48 Q^2)}{5 - 
      3 a^2 + 10 Q^2}\mathbf{g}\Bigr\}\,,\nn
\end{align}
where
$$\mathbf{g}=2 (5 Q^2 + 4) a_{-1}^6 + 45  a_{-1}^2 a_{-2}^2 - 42  a_{-3}^2
$$

\begin{align}
&Q_{\{1,3\},\{1,3\}}^{\mathrm{even}}(\{a_{-k}\})=
\frac1{1209600a^2(-21 (76 - 19 Q^2 - 30 Q^4) - (991 + 1076 Q^2) a^2 + 206 a^4)}
\nn\\&\times\Bigl\{
-a^2 \bigl[640 a^{10} (1 + 2 Q^2) - 
   16 a^8 (-27011 + 14098 Q^2 + 160 Q^4)  
   \nn\\&  +315 (1748 - 2969 Q^2 + 1830 Q^4) + 
   a^4 (6252242 - 9978784 Q^2 + 4263704 Q^4 - 2042880 Q^6)  \nn\\& + 
   4 a^6 (533225 - 1096594 Q^2 + 465312 Q^4 + 320 Q^6)   \nn\\&- 
   7 a^2 (-941629 + 942172 Q^2 - 466620 Q^4 + 102600 Q^6)\bigr]a_{-1}^8
  \nn\\& 
   +
   280 a^2 \bigl[-96 a^8 (1 + 2 Q^2) + 4 a^6 (-19249 + 9790 Q^2 + 96 Q^4)   \nn\\&-
    a^4 (279425 - 683886 Q^2 + 306784 Q^4 + 192 Q^6) + 
   315 (-380 - 82 Q^2 + 33 Q^4 + 450 Q^6)   \nn\\&+ 
   10 a^2 (-22552 + 129028 Q^2 - 60003 Q^4 + 32256 Q^6)\bigr]a_{-1}^4a_{-2}^2
   \nn\\&
   -420 a^2 \bigl[32 a^6 (1 + 2 Q^2) - 4 a^4 (-2217 + 5482 Q^2 + 32 Q^4)   \nn\\&+ 
   a^2 (49761 - 42996 Q^2 + 148256 Q^4 + 64 Q^6)   \nn\\&- 
   7 (-23408 - 36599 Q^2 + 22474 Q^4 + 19200 Q^6)\bigr]a_{-2}^4
   \nn\\&
   +
   3360 \bigl[16 a^6 (-8399 + 1742 Q^2) + 
   a^4 (473336 + 1270956 Q^2 - 338552 Q^4)   \nn\\&+ 
   a^2 (625801 - 664342 Q^2 - 3416448 Q^4 + 1059120 Q^6)  \nn\\& - 
   1575 (-76 - 401 Q^2 - 402 Q^4 + 372 Q^6 + 360 Q^8)\bigr]a_{-2}a_{-6}
   \nn\\&
   -5040 \bigl[32 a^6 (-2051 + 533 Q^2) + 
   a^4 (72508 + 457080 Q^2 - 124312 Q^4)  \nn\\& + 
   a^2 (85625 - 171248 Q^2 - 651588 Q^4 + 244656 Q^6)   \nn\\&- 
   315 (-76 - 401 Q^2 - 402 Q^4 + 372 Q^6 + 360 Q^8)\bigr]a_{-4}^2
\Bigr\}\,.\nn
\end{align}
\bigskip

{\it Acknowledgements.}\quad Research of SN is supported by Universit\`a Italo Francese
grant ``Vinci".
Research of FS is supported  by DIADEMS program (ANR) contract number BLAN012004.      
%%%%%%%%%%%%%%%%%%%%%%%%%%%%%%%%%%%%%%%%%%%%%%%%%%%%%%%%%%%%%%%%%%%%%%%%%%%%%%%%%%%%%%%%%%%%%%%%%%%%%%%%%%%%%%%%%%%%%%%%%%%%%%%%%
%%%%%%%%%%%

\end{document}